\documentclass[prl,twocolumn,showpacs,preprintnumbers,amsmath,amssymb]{revtex4}

\usepackage{graphics,epstopdf}
\usepackage{amsfonts}
\usepackage{amsthm}
\usepackage{comment}
\usepackage[usenames]{color}
\definecolor{AV}{rgb}{0.65,0.0,0}
\definecolor{GC}{rgb}{0,0.0,0.65}
\definecolor{WS}{rgb}{0,0.65,0}

\newcommand{\rom}[1]{{\mathrm{#1}}}

\usepackage[
      colorlinks=true,
      linkcolor=blue,
      urlcolor=blue,
      filecolor=blue,
      citecolor=red,
      pdfstartview=FitV,
      pdftitle={},
      pdfauthor={},
      pdfsubject={},
      pdfkeywords={},
      pdfpagemode=None,
      bookmarksopen=true
]{hyperref}

\usepackage{epsfig}

\def\d{\partial}

\def\eps{\epsilon}
\def\th{\theta}

\def\be{\begin{equation}}
\def\ee{\end{equation}}
\def\bea{\begin{eqnarray}}
\def\eea{\end{eqnarray}}

\begin{document}

\preprint{ULB-TH/10-32}

\title{Microscopics of Extremal Kerr from Spinning M5 Branes}

\author{Geoffrey Comp\`ere$^1$, Wei Song$^2$, and Amitabh Virmani$^3$ \\}
\address{$^1\,$
KdV Institute for Mathematics and Instituut voor Theoretische Fysica \\
Universiteit van Amsterdam, The Netherlands \\
$^2$\,Center for the Fundamental Laws of Nature, Harvard University, Cambridge, MA 02138, USA \\
 $^3\,$Physique Th\'eorique et Math\'ematique,  \\ Universit\'e Libre de
     Bruxelles and International Solvay Institutes, Bruxelles,  Belgium\\
{\tt \small gcompere@uva.nl, wsong@physics.harvard.edu, avirmani@ulb.ac.be} \\
}

\begin{abstract}
We show that the spinning
magnetic one-brane in minimal five-dimensional supergravity  admits a decoupling limit that interpolates
smoothly between a self-dual null orbifold of $AdS_3 \times S^2$ and the
near-horizon limit of the extremal Kerr black hole times a
circle. We use this interpolating solution to understand
the field theory dual to spinning M5 branes as a deformation of the
Discrete Light Cone Quantized (DLCQ) Maldacena-Stominger-Witten (MSW) CFT. In particular, the conformal weights of the
operators dual to the deformation around $AdS_3 \times S^2$ are calculated.
We present pieces of evidence showing that a CFT dual to the four-dimensional extremal Kerr can be obtained from the deformed MSW CFT.
\end{abstract}


\maketitle

Extremal rotating black holes lie in between the best understood but not realistic supersymmetric black holes and the less understood
but realistic neutral non-extremal black holes. For certain supersymmetric black holes, string theory has been used to identify the
microstates with those of two-dimensional conformal field theories (CFTs) \cite{Strominger:1996sh,Maldacena:1997de}. It has been conjectured that the extremal
Kerr black hole possesses a CFT description as well \cite{Guica:2008mu}. Recent progress has been made on the microscopic description of the  five-dimensional extremal Kerr-Newman black hole \cite{Guica:2010ej}\footnote{We thank A. Strominger and M. Guica for sharing drafts of \cite{Guica:2010ej} prior to publication.}. In this work, we aim at making progress in the microscopic description of the four-dimensional extremal Kerr black hole. Our construction shares many qualitative features with the one of \cite{Guica:2010ej} but differs significantly in details.

We proceed as follows. After reviewing the spinning magnetic one-brane in minimal
supergravity we discuss the near-horizon solution. Our crucial observation is that this solution smoothly
interpolates between the maximally supersymmetric self-dual null orbifold of $AdS_3 \times S^2$ and the near-horizon
limit of the extremal Kerr black hole times a circle. Using the
CFT description of the dual of string theory on $AdS_3 \times S^2$
 we infer that at zero angular momentum the dual description of the solution is in terms of the discrete light-cone quantized (DLCQ) Maldacena-Strominger-Witten (MSW) CFT \cite{Maldacena:1997de}.
Upon turning on the angular momentum, the CFT is deformed by a relevant and an irrelevant operator whose conformal weights and R-charges are identified. At small values of angular momentum, we observe that the generalized Kerr-CFT along the $z$ circle ($z$-CFT) has the same Virasoro generators and central charge as the deformed MSW CFT (DMSW). Therefore, we simply claim that DMSW is the microscopic definition of the $z$-CFT in the small angular momentum limit. The derivation of the finite deformations of the CFT is beyond the aspirations of this paper. We end with the observation that the central charge is not modified at leading order from the MSW CFT. We interpret this as an indication that the deformed theory is still a CFT. We provide one consistency numerical test and discuss the classical Virasoro symmetry implementing the conformal symmetry at finite deformation.

\section{Spinning M5 Branes}
Consider M-theory on $T^6 \times S^1$. Upon wrapping M5 branes appropriately on the internal 4-cycles of $T^6$, one obtains a charged string (one-brane) in the transverse five-dimensional space. The M5 charge is understood as a magnetic one-brane charge $Q_\rom{M} = \frac{1}{4\pi} \int_{S^2} F$ along the string. When an equal number of branes are wrapped along three four-tori of the $T^6$, the configuration is described in the low energy limit by five-dimensional minimal supergravity \cite{Chou:1997ba}.  The resulting finite temperature rotating black string was found in \cite{Compere:2009zh}. The solution depends on three parameters $(m,\beta,a)$. The five dimensional metric and gauge field take the form
\bea
ds^2\hspace{-4pt}& =&\hspace{-4pt} \bar h \left[ -  \frac{\Delta_2}{\xi} (dt + \omega_{\phi} d \phi)^2 +  \frac{dr^2}{\Delta}+d\theta^2 + \frac{\Delta}{\Delta_2}\sin^2\theta d\phi^2 \right]\nonumber\\
& +& \xi \bar h^{-2} \left( dz + \hat A_{t} dt + \hat A_{\phi} d \phi \right)^2, \quad A_t = -\sqrt{3}F_3 \bar h^{-1}\nonumber\\
A_z &=& -\sqrt{3} \frac{F_2}{ \bar h} ,\;\;  A_\phi = -\sqrt{3}F_0 \frac{\Delta}{\Delta_2} + \bar A_\phi A_z + \omega_\phi A_t\label{magnetic}
\eea
where we have defined the following functions
\bea
\hat A_t  &=& - \frac{F_2 F_3}{\xi},\; \hat A_\phi = \bar A_\phi + \hat A_t \omega_\phi,\;  \xi = \bar h (F_4+\Delta_2) -F_2^2\nonumber\\
\omega_\phi &=& \frac{2a m c^3}{\Delta_2}r \sin^2\theta,\quad \bar A_\phi = \frac{2a m s^3}{\Delta_2}(r-2m)\sin^2\theta,\nonumber\\
\bar h &=& \Delta_2 + 2 m (r(c^2-s^2)+2m s^4),\;\; F_0 = 2 m c s \cos\theta\nonumber\\
F_2 &=& a s F_0, \quad F_3 = -a c F_0,\quad F_4 = 2mc^2 r \nonumber \eea
\bea
\Delta &=& r^2 -2mr+a^2,\quad \Delta_2 = \Delta -a^2 \sin^2\theta
\eea
and we have introduced $s := \sinh \beta,\; c := \cosh \beta$. The orientation is $\epsilon_{trz\theta\phi}=-\sqrt{-g}$. The extremal solution $m = a$ has zero Hawking temperature but is not supersymmetric for $a > 0$. For $\beta = 0$, the solution reduces to extremal Kerr black hole times a circle. When $a>0$ there is a horizon at $r_+ = a$ but when $a = 0$ the solution reduces to flat space $\mathbb R^{3,1} \times S^1$.

The mass, magnetic charge, angular momentum and entropy of the extremal solution from the five-dimensional perspective are
\bea
M ={\pi L_z a(2+3s^2)\over G_5}, &\quad& Q_M = 2\sqrt{3} a s c, \\ J_\phi = \frac{2 \pi L_z a^2 c^3}{G_5}, &\quad& S = 2 \pi J_\phi.
\eea
The five-dimensional Newton's constant is related to the four-dimensional and eleven-dimensional Newton's constants as  $ G_5 = (2\pi L_z) G_4 = G_{11}/(V_6)$, where $L_z$ is the length of the circle $S^1$ and $V_6$ is the volume of the six-torus $T^6$.  We introduce the number of free M5 branes as
\be
n = \frac{2 Q_\rom{M}}{ \sqrt{3} l_p} = \frac{2a \sinh (2\beta)}{l_p},\label{nbeta}
\ee
where $l_p = (4 G_5/\pi)^{1/3}$ is the five-dimensional Planck length. Both the angular momentum $J_\phi$ and the number of M5 branes are expected to be quantized as $n \in \mathbb Z$, $J_\phi \in \mathbb Z/2$.

The angular and linear velocity at the horizon are
\be
\Omega_\phi = \frac{1}{2 a \cosh^3\beta}, \quad
v_z = - \tanh^3\beta.
\ee
Note that even though the ADM linear momentum is zero, the solution exhibits frame dragging along the string direction near the horizon due to the linear velocity  \cite{Compere:2009zh}. This indicates that there is an interesting dynamics close to the horizon to which we now turn our attention.


\section{ Near Horizon Limit}
Let us take a near-horizon limit of the solution \eqref{magnetic} in comoving coordinates
\be
r \rightarrow a+ \mu r, \; t \rightarrow \frac{t}{\mu}, \; \phi \rightarrow \phi + \Omega_\phi \frac{t}{\mu}, \; z \rightarrow z + v_z \frac{t}{\mu}
\ee
by sending $\mu \rightarrow 0$. This limit is a decoupling limit where the asymptotically flat region is no longer part of the spacetime. It is also convenient to rescale the coordinates and the parameter $a$ as
\be
t \rightarrow l_p^{2}\hat R^2  t/2 \qquad a \rightarrow l_p \hat a, \qquad z \rightarrow l_p \hat R z\label{dec2}.
\ee
where $\hat R$ is a constant. We denote by $2\pi \hat L_z = 2\pi L_z/(\hat R \, l_p)$ the length of the circle along the final $z$ coordinate.

The resulting geometry and gauge field is a solution of minimal supergravity. The geometry has enhanced isometry $SL(2,\mathbb{R})\times U(1)\times U(1)$, as familiar from the attractor mechanism \cite{Astefanesei:2006dd, Astefanesei:2007bf} and general near-horizon limits \cite{Kunduri:2007vf}. The solution reads as
\begin{eqnarray}
\frac{ds^2}{R^2\, l_p^2} &=& \Gamma(\theta)\left[-(k_\phi)^2 r^2dt^2+\frac{dr^2}{r^2} + d\theta^2 \right]\nonumber \\
&&+ \gamma_{\phi \phi}(\theta)e_\phi^2+ 2\gamma_{\phi z}(\theta) e_\phi \, e_z + \gamma_{zz}(\theta) e_z^2 \nonumber \\
\frac{A}{R \, l_p} &=& f_\phi(\theta)e_\phi  +f_z(\theta) e_z  \ ,  \label{GenExt}
\end{eqnarray}
where $e_\phi = d\phi + k_\phi r dt$, $e_z = dz + k_z r dt$ and
all functions are more easily expressed in terms of the variables $R > 0$ and $\Phi \in [0,\frac{\pi}{2}]$ defined by
\be
n = R \cos\Phi \equiv R c_\Phi, \qquad \hat a = \frac{1}{2} R \sin \Phi \equiv  \frac{1}{2} R s_\Phi .
\ee

When one chooses $\hat R = R$ in \eqref{dec2}, one finds that all functions only depend on $\Phi$ as
\bea
\Gamma(\theta) &=& {1\over4}{(1+ s^2_\Phi c^2_\theta)}, \quad f_\phi(\th) = \frac{\sqrt{3}} {8\Gamma(\theta)} c_\theta c_\Phi (c^2_\Phi-2),\nonumber\\
k_\phi &=&  \frac{8(1+c_\Phi)^{3/2}}{s^{1/2}_\Phi (1+c_\Phi +s_\Phi)^3}, \; k_z = -\frac{2 (1-s_\Phi+c_\Phi)^3}{s_\Phi (1+s_\Phi+c_\Phi)^3},\nonumber\\
f_z(\th) &=& -\frac{\sqrt{3} c_\theta (-s_\Phi+c_\Phi+1) s^{1/2}_\Phi c_\Phi}{8\Gamma(\theta) (c_\Phi+1)^{1/2} },\\
\gamma_{\phi \phi}(\th ) &=&  \left(64\Gamma(\theta)^2\right)^{-1} \sin^2\theta [s_\Phi^4(3+s_\Phi^2)c^2_\theta+3s_\Phi^2+1],\nonumber \\
\gamma_{\phi z} (\th)&=&-\left(32 k_\phi\Gamma(\theta)^2\right)^{-1}s_\Phi  k_z \sin ^2\theta  \left(1-s_\Phi^3 c^2_\theta\right),\nonumber\\
\gamma_{zz} (\th)&=& \frac{s_\Phi \left( 2s_\Phi^3 c^4_\theta - (1+s_\Phi)(s_\Phi^2-4s_\Phi+1)c^2_\theta+2\right) }{32
\Gamma(\theta)^2}, \nonumber
\eea
where $c_\theta \equiv \cos\theta$. 
Note that $k_\phi$ and $k_z$ diverge as $\hat a$ approaches zero.


\section{Interpolating Geometry}

We now present two limits of the decoupled solution \eqref{GenExt} when either the number of M5 branes $n$ or the angular momentum $J_\phi$ is zero. When there is no M5 brane charge ($n = 0$ or $\Phi = \frac{\pi}{2}$), the solution \eqref{GenExt} reduces to the NHEK geometry \cite{Bardeen:1999px} times a circle
\bea
\frac{ds^2}{4 G_4 J_\phi} &=& \Gamma(\theta) \left( \frac{dr^2}{r^2} +d\theta^2 - r^2 dt^2 \right) \\  & &  + \gamma_{\phi\phi}(\theta) (d\phi + r dt)^2\nonumber
+ dz^2, \: \: A = 0,  
\eea
where
\be
\Gamma(\theta) = \frac{1}{4}(1+\cos^2\theta),~ \gamma_{\phi\phi}(\theta) = \frac{ \sin^2\theta}{1+\cos^2\theta}\,.\ee
This is expected since the original extremal spinning one-brane solution reduces to the extremal Kerr spacetime times a circle.

Now, the crucial observation is that the decoupled solution \eqref{GenExt} is smooth upon setting $J_\phi =0$ at fixed brane number $n=R$ (equivalently,  $\Phi =0$, or $a \rightarrow 0$, $\beta \rightarrow \infty$ and $n$ fixed in the original parameterization \eqref{nbeta}). This decoupled solution for $\Phi = 0$ is just the null orbifold of $AdS_3$ times $S^2$,
\bea
ds^2 &=& l^2 \left( \frac{dr^2}{4r^2} - 2 r  dt dz \right) +l^2_{S_2}d\Omega_2 ,\nonumber \\
A &=& -\sqrt{3} l_{S_2}\cos\theta d\phi, \quad z \sim z + 2\pi \hat L_z, \label{ads3geom}
\eea
where the two sphere has radius $l_{S_2} = \frac{1}{2} R\, l_p$ half of the $AdS_3$ radius $l = R\, l_p$.
The solution has zero entropy and angular momentum.
The local isometry group is  $SL(2,\mathbb R)_L\times SL(2,\mathbb R)_R \times SO(3)_R$, with the $SL(2,\mathbb R)_R$ generators
\bea
\bar H_{-1}&=&\partial_t\qquad \bar H_{0}=-t\partial_t+r\partial_r\\
\bar H_{1}&=&t^2\partial_t-2tr\partial_r+{1\over2r}\partial_z \label{unbr}
\eea
and with the $SL(2,\mathbb R)_L$ generators $H_{n}$, $n=-1,0,1$ obtained by exchanging $t\leftrightarrow z$.
The null identification $z \sim z + 2\pi \hat L_z$  breaks the $SL(2,\mathbb R)_L$ to $U(1)_L$ generated by $\d_z.$

The isometry supergroup of $AdS_3 \times S^2$ is $SU(1,1 | 2)_R \times
SL(2,\mathbb R)_L$.  This can be checked by explicitly constructing
Killing spinors. The Killing spinors for \eqref{ads3geom} (without the
null identification) are given in \cite{Gauntlett:2002nw}. The null
identification does not bring in any change; and the geometry
\eqref{ads3geom} is maximally supersymmetric admitting $8$
$z$-periodic Killing spinors. For $J_\phi > 0$, supersymmetry is
completely broken. This is because the existence
of a Killing spinor implies the existence of a global
timelike or null Killing vector \cite{Gauntlett:2002nw}, but there is no such vector
in the decoupled geometry (8) at $J_\phi > 0$. This is consistent with the fact that the BPS bound
\be
\frac{4G_4 M}{\sqrt{3} Q_M} = \sec\Phi+\frac{1}{3}\tan\Phi \geq 1 \label{BPS}
\ee
is saturated if and only if $\Phi=0$.

In summary, one can understand the decoupled solution as a solution
interpolating smoothly between the near horizon limit of the extremal
Kerr black hole times a 
circle and the  supersymmetric null orbifold of
$AdS_3$ times a two-sphere.

\section{AdS$_3$, Temperature and DLCQ}
String theory on $AdS_3 \times S^2$ is dual to the MSW CFT
\cite{Maldacena:1997de} \footnote{Strictly speaking this is only true for the case of M-theory on a Calabi-Yau manifold \cite{Maldacena:1997de}. In this case the dual CFT is the worldvolume theory of M5 branes, and it is often referred to as the MSW CFT. For simplicity, we continue to refer the CFT dual to our set up also as the MSW CFT.}. The null orbifold \eqref{ads3geom} has its
dual description in terms of discrete light-cone quantized (DLCQ)
MSW CFT \cite{Seiberg:1997ad, Balasubramanian:2009bg}.

To see this, it is more convenient to redefine
$t'=k_\phi t$.
Let us fix $R$ and consider $\hat a = R \epsilon^2$ where $\epsilon$ is a small expansion parameter.
We find that up to terms $O(\eps^4)$, the metric can be written as
\bea
ds^2&=&ds^2_3+l^2_{S_2}(dy^i-A^{ij}y^j)^2,\,\label{kkRedu}\\
\frac{4 ds^2_3}{l^2}&=&-(1_\eps)^2 r^2dt'^2+{dr^2\over {r^2} }
+\left(2\epsilon dz-1_\eps rdt'\right)^2, \label{ads3eps2}
\eea
where $y^i$ with $i=1,2,3,~\sum(y^i)^2=1$ are coordinates on the unit two-sphere, $1_\eps \equiv 1+3\eps^2$ and $z \sim z+2\pi \hat{L}_z$. The SU(2) gauge fields $A^{ij}=-A^{ji}$ are presented  below. The three dimensional metric \eqref{ads3eps2} is a spacelike quotient of $AdS_3$. The quotient is defined by the group element $e^{i2\pi \hat{L}_z\d_ z}$.  Then $\hat{L}_z\d_ z \equiv \pi T_L H_L $, where $H_L$ is a unit normalized $SL(2,\mathbb R)_L$ generator in global $AdS_3$, and $T_L$ is defined as a dimensionless left moving temperature \cite{Maldacena:1998bw}. One finds \be T_L={\epsilon\hat{L}_z\over {\pi }}+O(\eps^3).\label{TL}\ee
To get back to \eqref{ads3geom}, we need to rescale back using $t'=2t/( 1_\epsilon\epsilon)$ and take the limit $\epsilon \to 0$. In this limit the  right-moving sector freezes due to the infinite redshift in $t$. The allowed excitations are only in the left-moving sector. One thus obtains the DLCQ MSW CFT. Also note that both $T_L$ and the entropy $S$ approach zero in this limit. The Virasoro generators reproducing Cardy formula in this limit are
\be
L^z_n = \hat L_z e^{i n z / \hat L_z} \partial_z - i n  e^{i n z / \hat L_z} r \partial_r + \dots +O(\eps^2) \label{kerrCFTz3}\\
\ee
where $\dots$ indicate subleading terms in $r$. The Brown-Henneaux central charge of $AdS_3$ is $c_L = \frac{3 l}{2 G_3}+O(\eps^2) = 6 R^3 + O(\eps^2)$. The entropy is indeed reproduced by Cardy's formula $S_\rom{micro}={\pi^2\over3}c_L T_L = 2\pi J_\phi+O(\eps^3)$.

We conclude that the DLCQ MSW CFT with $c = 6 R^3$ controls the physics of string theory on the interpolating solution in the limit $J_\phi \rightarrow 0$.

\section{Operator Deformations}
Let us now consider the U(1) gauge field \eqref{GenExt} and the SU(2) gauge fields \eqref{kkRedu}, and keep them up to order $\epsilon^3$ in the $t'$ coordinates. Part of this analysis is inspired by \cite{Guica:2010}. Choosing $y^1=\sin\theta\cos\phi,~y^2=\sin\theta\sin\phi,~y^3=\cos\theta,$ we find
\bea  A^{12}&=&2\epsilon\left((1-3\epsilon^2)dz+3\epsilon rdt'\right),\\
A&=&-\frac{\sqrt{3}}{2}l  y^3 \left(d\phi +2\epsilon  \left((1-\epsilon^2)dz+2\epsilon rdt'\right)\right). \label{gauge2}\eea
The perturbation can be written as the sum $(\delta A^{ij},\delta A,\delta ds^2_3) = \eps M_-+ \eps^2 M_+ - \eps^3 M_- +N.L.$ where the two linear modes are
\bea
M_+ &=& (\delta A^{12} =6 r dt',\; \; {\delta A\over l y^3} = -2\sqrt{3}  r dt',\;\; \delta ds_3^2=0),\nonumber\\
M_- &=& (\delta A^{12}= 2dz, \;\; {\delta A\over l y^3} = -\sqrt{3}dz, \;\; \delta ds_3^2= 0),
\eea
and the non-linear terms are
\bea
N.L. =  (\delta A^{12}= -4 \eps^3 dz, \;\; {\delta A\over l y^3} = 0, \;\; \delta ds_3^2= \eps^2 l^2 dz^2).
\eea
Note the following properties of these linear modes under the conformal group
\be
\mathcal L_{\bar H_0} M_\pm = 0, \qquad \mathcal L_{H_0} M_\pm = \pm M_\pm .
\ee
The first property is a consequence of the invariance of the perturbation under the rescaling $r \rightarrow \lambda r$, $t' \rightarrow t'/\lambda$. It implies that the dual operator sourced by $M_\pm$ has right-moving conformal weight $h_R = 1$. The second property implies that the left-moving conformal weights are $h_L = 1 \pm 1$. Since these perturbations are both expressed in terms of the $k=1$ vector harmonics upon dimensional reduction on the sphere, the R-charge is $q=k=1$. In summary, the perturbation are sources to operators with conformal weights $(h_L,h_R,q) = (2,1,1)$ for $M_+$ and $(h_L,h_R,q) = (0,1,1)$ for $M_-$. Their spin is respectively $s \equiv h_R - h_L = \mp 1$.

In terms of the representations of $SU(1,1|2)_R\times SL(2,\mathbb{R})_L$, the mode $M_+$ corresponds to a chiral primary operator of a vector multiplet while $M_-$ corresponds to a chiral primary operator of a graviton multiplet \cite{Fujii:1998tc}.  In both cases they lie in the so-called `quarteton' representation.

We note that the $M_-$ perturbation is generated by the large coordinate transformation $\phi \rightarrow \phi + 2\epsilon z $. Therefore, the $(0,1,1)$ operator can be recognized as an operator generating a spectral flow along a $U(1) \subset SU(2)_R$ \cite{deBoer:2008fk}. The nature of the irrelevant operator $(2,1,1)$ remains to be investigated.

\section{From MSW CFT to Kerr-CFT at small $J_\phi$}
 Up to order $\epsilon^3$ in the metric, we saw that there are two consequences of turning on the angular momentum. First, it changes the three dimensional metric to \eqref{ads3eps2}, which amounts to turning on a left moving temperature \eqref{TL}. Second, it changes the gauge fields \eqref{gauge2}, which amounts to deforming the DLCQ MSW CFT by two operators.  Supersymmetry is broken at second order in $\eps$ as can be seen from \eqref{BPS} with $\Phi =2\eps^2$. By analyzing the operator deformation, one should be able to know the nature of the deformed CFT. We leave a systematic investigation for the future. Let us call the deformed theory DMSW.
At this order in the deformation, we have a set of Virasoro generators \eqref{kerrCFTz3} with the central charge $c = 6 R^3 + O(\eps^2)$, and the Cardy formula holds. Therefore, DMSW is still a CFT at least in the classical limit.
On the other hand, boundary analysis in the spirit of Kerr-CFT \cite{Brown:1986ed,Guica:2008mu} predicts the existence of a boundary CFT. In particular, we can define Virasoro generators along the $z$-circle, which are the same as \eqref{kerrCFTz3} without the $O(\epsilon^2)$ terms. The central charge and temperature are $c_z=6n^3=6R^3+O(\epsilon^4), T_z={J_\phi\over\pi n^3}={\epsilon\hat{L}_z\over {\pi }}+O(\eps^3)$. Up to this order, we see that DMSW and the $z$-CFT have the same known properties. We therefore simply claim DMSW as the microscopic definition of $z$-CFT as $\epsilon\rightarrow 0$.

\section{Extrapolation to finite deformation}
Starting from $\epsilon^4$ the three dimensional metric \eqref{ads3eps2} is no longer locally $AdS_3$. Without knowing the effect of the operator deformations of the CFT, we cannot say anything conclusive. Let us propose, however, in line with the Kerr-CFT conjecture \cite{Guica:2008mu}, that the DMSW is still a CFT at all orders of $\eps$.
Moreover, let us extrapolate the fact that the central charge is not modified by linear $\eps$ terms, into the conjecture that the classical central charge of the DMSW is exactly $c_{D}= 6 R^3$ at all orders in $\eps$, so that we have
\be
c_{D} = 6 R^3 ,\quad T_{D} = \frac{J_\phi}{\pi R^3},\quad S_{\rom{micro}} = \frac{\pi^2}{3}c_{D} T_{D}.
\label{dMSW}
\ee
We now present a consistency check for this conjecture and discuss classical Virasoro algebras in the bulk that reproduce \eqref{dMSW}.

The brane charge $n$ is integer quantized. At $J_\phi=0$, $R=n$, so $R$ is integer quantized as well. Moreover, the angular momentum $J_\phi$ should be half-integer quantized. We consider the DMSW at fixed central charge $6R^3$ but we allow $n$ and $J_\phi$ to vary. In the classical limit, $n$ and $J_\phi$ are both associated with the angle $\Phi$ ranging from 0 to $\pi/2$. The angular momentum at $\Phi = \pi/2$ is maximal and given by $J_\phi^\rom{max} = 2 R^3 \hat L_z$. Let us fix $\hat L_z$ such that $J_\phi^\rom{max}$ is a fixed half-integer. In order to have a non-trivial theory, there must exist half-integer intermediate angular momenta with $0 \leq J_\phi \leq J_\phi^\rom{max}$, which correspond classically to geometries with $0 < \Phi < \pi/2$. Using the classical formula for the angular momentum, they have to obey
\be
\frac{J_\phi}{J_\phi^\rom{max}}  = \left( 1-n^2/R^2 \right)^{1/4} \left(\frac{1}{2} + \frac{1}{2} (1-n^2/R^2)^{1/2} \right)^{3/2}\label{quantization1} \, .
\ee
There are clearly a large number of solutions to this equation. For example, $R = (r^2+s^2)^2+4r^2 s^2$, $n = 4 r s (r^2+s^2)$, $J_\phi^\rom{max} = t R^2/2$ and $J_\phi = (r^2+s^2)^3(r^2-s^2)t/2$ are all solutions for $r$, $s$, $t$ taking any integer values.

There are many sets of Virasoro generators and associated boundary conditions in the decoupled geometry such that the Virasoro algebra has central charge $6R^3$. One can perform the $SL(2,\mathbb Q)$ transformation $u={1\over q}({p_1}\phi-{p_2}{z\over\hat{L}_z})$, $v={1\over q}(-{p_3}\phi+{p_4}{z\over\hat{L}_z})$ where $p_1 p_4 -p_2 p_3=q^2$ and consider the following vector fields
\be L_n^u=f_n(u)\d_u-f'_n(u)r\d_r,\qquad f_n(u)=-e^{-inu}.~ \label{Viru}\ee
In order for \eqref{Viru} to be single valued, we need to restrict to the subset where $ n=k q,\; k\in \mathbb Z. $
When $q=1$, one can choose the boundary conditions to make \eqref{Viru} the Virasoro generators of the asymptotic symmetry group \cite{Azeyanagi:2008kb,Chen:2009xja}. For some subtleties in the definition of charges, see also \cite{Amsel:2009pu}.
When $q>1$, it can be understood as a long string picture of the $q=1$ case \cite{Maldacena:1996ds,Guica:2010ej}.
 The central charge for $u=\phi$ is $c_\phi=12 J_\phi,$ while for $u=-{z\over\hat{L}_z}$ is $c_z=6n^3$. More generally, $c_u={1\over q}({p_1}c_\phi+{p_2}c_z)$.
 Given the charges satisfying \eqref{quantization1}, there are always integers $q,p_1,p_2$ satisfying the equation \be R^3q=2J_\phi p_1+n^3p_2.\ee Plugging these solutions into \eqref{Viru}, and choosing boundary conditions correspondingly, we get a family of Virasoro generators with the central charge $c_u=6R^3,$ which define a family of CFTs labeled by $(q, p_1,p_2)$.
 The corresponding temperature can be obtained by analyzing the Frolov-Thorne vacuum, and it is $T_u=\frac{J_\phi}{\pi R^3}$. If DMSW exists and has the property \eqref{dMSW}, then we should be able to identify it with one particular member of the family of $(q, p_1,p_2)$ Kerr-CFTs \footnote{This statement could be modified if we take possible current algebras into account.}. As $J_\phi\rightarrow 0$, the Virasoro generators and the central charge of the DMSW agree with that of the (generalized) Kerr-CFT along the $z$ circle, or equivalently, $(q, p_1,p_2)$ Kerr-CFT with $p_1\rightarrow0, {p_2\over q}\rightarrow 1$. Therefore, we simply claim that the DMSW CFT is the microscopic definition of the Kerr-CFT along the $z$ circle in the small angular momentum limit. As $n\rightarrow 0,$ the original Kerr-CFT analysis in four dimensions shows that the Virasoro generators are parameterized by $\phi$. Thus, we expect that the DMSW has $p_2=0$ in this limit and is related to the four dimensional Kerr-CFT by going to the long or short string picture. It remains a puzzle that extra considerations are needed to  uniquely fix the Virasoro algebra.


\section*{Acknowledgments}
We thank R. Benichou, J. de Boer, M. Guica, T. Hartman, G. Horowitz, S.-S. Kim, D. Marolf, J. Simon and A. Strominger for stimulating discussions.
AV thanks the organizers of Monte Verita Conference on Strings, M-theory and Quantum Gravity for their kind hospitality.
GC thanks Harvard University and UCSB for their kind hospitality and he acknowledges NWO through a VICI grant.
WS was supported by the Harvard Society of Fellows and DOE grant DEFG0291ER40654.
AV was supported by IISN Belgium (conventions 4.4511.06 and 4.4514.08), and by the Belgian Federal Science Policy Office through the Interuniversity Attraction Pole P6/11.

\bibliography{master4}

\begin{thebibliography}{22}
\expandafter\ifx\csname natexlab\endcsname\relax\def\natexlab#1{#1}\fi
\expandafter\ifx\csname bibnamefont\endcsname\relax
  \def\bibnamefont#1{#1}\fi
\expandafter\ifx\csname bibfnamefont\endcsname\relax
  \def\bibfnamefont#1{#1}\fi
\expandafter\ifx\csname citenamefont\endcsname\relax
  \def\citenamefont#1{#1}\fi
\expandafter\ifx\csname url\endcsname\relax
  \def\url#1{\texttt{#1}}\fi
\expandafter\ifx\csname urlprefix\endcsname\relax\def\urlprefix{URL }\fi
\providecommand{\bibinfo}[2]{#2}
\providecommand{\eprint}[2][]{\url{#2}}

\bibitem[{\citenamefont{Strominger and Vafa}(1996)}]{Strominger:1996sh}
\bibinfo{author}{\bibfnamefont{A.}~\bibnamefont{Strominger}} \bibnamefont{and}
  \bibinfo{author}{\bibfnamefont{C.}~\bibnamefont{Vafa}},
  \bibinfo{journal}{Phys. Lett.} \textbf{\bibinfo{volume}{B379}},
  \bibinfo{pages}{99} (\bibinfo{year}{1996}), \eprint{hep-th/9601029}.

\bibitem[{\citenamefont{Maldacena et~al.}(1997)\citenamefont{Maldacena,
  Strominger, and Witten}}]{Maldacena:1997de}
\bibinfo{author}{\bibfnamefont{J.~M.} \bibnamefont{Maldacena}},
  \bibinfo{author}{\bibfnamefont{A.}~\bibnamefont{Strominger}},
  \bibnamefont{and} \bibinfo{author}{\bibfnamefont{E.}~\bibnamefont{Witten}},
  \bibinfo{journal}{JHEP} \textbf{\bibinfo{volume}{12}}, \bibinfo{pages}{002}
  (\bibinfo{year}{1997}), \eprint{hep-th/9711053}.

\bibitem[{\citenamefont{Guica et~al.}(2009)\citenamefont{Guica, Hartman, Song,
  and Strominger}}]{Guica:2008mu}
\bibinfo{author}{\bibfnamefont{M.}~\bibnamefont{Guica}},
  \bibinfo{author}{\bibfnamefont{T.}~\bibnamefont{Hartman}},
  \bibinfo{author}{\bibfnamefont{W.}~\bibnamefont{Song}}, \bibnamefont{and}
  \bibinfo{author}{\bibfnamefont{A.}~\bibnamefont{Strominger}},
  \bibinfo{journal}{Phys. Rev.} \textbf{\bibinfo{volume}{D80}},
  \bibinfo{pages}{124008} (\bibinfo{year}{2009}), \eprint{0809.4266}.

\bibitem[{\citenamefont{Guica and Strominger}(2010)}]{Guica:2010ej}
\bibinfo{author}{\bibfnamefont{M.}~\bibnamefont{Guica}} \bibnamefont{and}
  \bibinfo{author}{\bibfnamefont{A.}~\bibnamefont{Strominger}}
  (\bibinfo{year}{2010}), \eprint{1009.5039}.

\bibitem[{\citenamefont{Chou et~al.}(1997)}]{Chou:1997ba}
\bibinfo{author}{\bibfnamefont{A.~S.} \bibnamefont{Chou}} \bibnamefont{et~al.},
  \bibinfo{journal}{Nucl. Phys.} \textbf{\bibinfo{volume}{B508}},
  \bibinfo{pages}{147} (\bibinfo{year}{1997}), \eprint{hep-th/9704142}.

\bibitem[{\citenamefont{Compere et~al.}(2009)\citenamefont{Compere, de~Buyl,
  Jamsin, and Virmani}}]{Compere:2009zh}
\bibinfo{author}{\bibfnamefont{G.}~\bibnamefont{Compere}},
  \bibinfo{author}{\bibfnamefont{S.}~\bibnamefont{de~Buyl}},
  \bibinfo{author}{\bibfnamefont{E.}~\bibnamefont{Jamsin}}, \bibnamefont{and}
  \bibinfo{author}{\bibfnamefont{A.}~\bibnamefont{Virmani}},
  \bibinfo{journal}{Class. Quant. Grav.} \textbf{\bibinfo{volume}{26}},
  \bibinfo{pages}{125016} (\bibinfo{year}{2009}), \eprint{0903.1645}.

\bibitem[{\citenamefont{Astefanesei et~al.}(2006)\citenamefont{Astefanesei,
  Goldstein, Jena, Sen, and Trivedi}}]{Astefanesei:2006dd}
\bibinfo{author}{\bibfnamefont{D.}~\bibnamefont{Astefanesei}},
  \bibinfo{author}{\bibfnamefont{K.}~\bibnamefont{Goldstein}},
  \bibinfo{author}{\bibfnamefont{R.~P.} \bibnamefont{Jena}},
  \bibinfo{author}{\bibfnamefont{A.}~\bibnamefont{Sen}}, \bibnamefont{and}
  \bibinfo{author}{\bibfnamefont{S.~P.} \bibnamefont{Trivedi}},
  \bibinfo{journal}{JHEP} \textbf{\bibinfo{volume}{10}}, \bibinfo{pages}{058}
  (\bibinfo{year}{2006}), \eprint{hep-th/0606244}.

\bibitem[{\citenamefont{Astefanesei and Yavartanoo}(2008)}]{Astefanesei:2007bf}
\bibinfo{author}{\bibfnamefont{D.}~\bibnamefont{Astefanesei}} \bibnamefont{and}
  \bibinfo{author}{\bibfnamefont{H.}~\bibnamefont{Yavartanoo}},
  \bibinfo{journal}{Nucl. Phys.} \textbf{\bibinfo{volume}{B794}},
  \bibinfo{pages}{13} (\bibinfo{year}{2008}), \eprint{0706.1847}.

\bibitem[{\citenamefont{Kunduri et~al.}(2007)\citenamefont{Kunduri, Lucietti,
  and Reall}}]{Kunduri:2007vf}
\bibinfo{author}{\bibfnamefont{H.~K.} \bibnamefont{Kunduri}},
  \bibinfo{author}{\bibfnamefont{J.}~\bibnamefont{Lucietti}}, \bibnamefont{and}
  \bibinfo{author}{\bibfnamefont{H.~S.} \bibnamefont{Reall}},
  \bibinfo{journal}{Class. Quant. Grav.} \textbf{\bibinfo{volume}{24}},
  \bibinfo{pages}{4169} (\bibinfo{year}{2007}), \eprint{0705.4214}.

\bibitem[{\citenamefont{Bardeen and Horowitz}(1999)}]{Bardeen:1999px}
\bibinfo{author}{\bibfnamefont{J.~M.} \bibnamefont{Bardeen}} \bibnamefont{and}
  \bibinfo{author}{\bibfnamefont{G.~T.} \bibnamefont{Horowitz}},
  \bibinfo{journal}{Phys. Rev.} \textbf{\bibinfo{volume}{D60}},
  \bibinfo{pages}{104030} (\bibinfo{year}{1999}), \eprint{hep-th/9905099}.

\bibitem[{\citenamefont{Gauntlett et~al.}(2003)\citenamefont{Gauntlett,
  Gutowski, Hull, Pakis, and Reall}}]{Gauntlett:2002nw}
\bibinfo{author}{\bibfnamefont{J.~P.} \bibnamefont{Gauntlett}},
  \bibinfo{author}{\bibfnamefont{J.~B.} \bibnamefont{Gutowski}},
  \bibinfo{author}{\bibfnamefont{C.~M.} \bibnamefont{Hull}},
  \bibinfo{author}{\bibfnamefont{S.}~\bibnamefont{Pakis}}, \bibnamefont{and}
  \bibinfo{author}{\bibfnamefont{H.~S.} \bibnamefont{Reall}},
  \bibinfo{journal}{Class. Quant. Grav.} \textbf{\bibinfo{volume}{20}},
  \bibinfo{pages}{4587} (\bibinfo{year}{2003}), \eprint{hep-th/0209114}.

\bibitem[{\citenamefont{Seiberg}(1997)}]{Seiberg:1997ad}
\bibinfo{author}{\bibfnamefont{N.}~\bibnamefont{Seiberg}},
  \bibinfo{journal}{Phys. Rev. Lett.} \textbf{\bibinfo{volume}{79}},
  \bibinfo{pages}{3577} (\bibinfo{year}{1997}), \eprint{hep-th/9710009}.

\bibitem[{\citenamefont{Balasubramanian
  et~al.}(2010)\citenamefont{Balasubramanian, de~Boer, Sheikh-Jabbari, and
  Simon}}]{Balasubramanian:2009bg}
\bibinfo{author}{\bibfnamefont{V.}~\bibnamefont{Balasubramanian}},
  \bibinfo{author}{\bibfnamefont{J.}~\bibnamefont{de~Boer}},
  \bibinfo{author}{\bibfnamefont{M.~M.} \bibnamefont{Sheikh-Jabbari}},
  \bibnamefont{and} \bibinfo{author}{\bibfnamefont{J.}~\bibnamefont{Simon}},
  \bibinfo{journal}{JHEP} \textbf{\bibinfo{volume}{02}}, \bibinfo{pages}{017}
  (\bibinfo{year}{2010}), \eprint{0906.3272}.

\bibitem[{\citenamefont{Maldacena and Strominger}(1998)}]{Maldacena:1998bw}
\bibinfo{author}{\bibfnamefont{J.~M.} \bibnamefont{Maldacena}}
  \bibnamefont{and}
  \bibinfo{author}{\bibfnamefont{A.}~\bibnamefont{Strominger}},
  \bibinfo{journal}{JHEP} \textbf{\bibinfo{volume}{12}}, \bibinfo{pages}{005}
  (\bibinfo{year}{1998}), \eprint{hep-th/9804085}.

\bibitem[{\citenamefont{Guica et~al.}(201x)\citenamefont{Guica, Hartman, Song,
  and Strominger}}]{Guica:2010}
\bibinfo{author}{\bibfnamefont{M.}~\bibnamefont{Guica}},
  \bibinfo{author}{\bibfnamefont{T.}~\bibnamefont{Hartman}},
  \bibinfo{author}{\bibfnamefont{W.}~\bibnamefont{Song}}, \bibnamefont{and}
  \bibinfo{author}{\bibfnamefont{A.}~\bibnamefont{Strominger}},
  \bibinfo{journal}{in progress}  (\bibinfo{year}{201x}).

\bibitem[{\citenamefont{Fujii et~al.}(2000)\citenamefont{Fujii, Kemmoku, and
  Mizoguchi}}]{Fujii:1998tc}
\bibinfo{author}{\bibfnamefont{A.}~\bibnamefont{Fujii}},
  \bibinfo{author}{\bibfnamefont{R.}~\bibnamefont{Kemmoku}}, \bibnamefont{and}
  \bibinfo{author}{\bibfnamefont{S.}~\bibnamefont{Mizoguchi}},
  \bibinfo{journal}{Nucl. Phys.} \textbf{\bibinfo{volume}{B574}},
  \bibinfo{pages}{691} (\bibinfo{year}{2000}), \eprint{hep-th/9811147}.

\bibitem[{\citenamefont{de~Boer et~al.}(2008)\citenamefont{de~Boer, Denef,
  El-Showk, Messamah, and Van~den Bleeken}}]{deBoer:2008fk}
\bibinfo{author}{\bibfnamefont{J.}~\bibnamefont{de~Boer}},
  \bibinfo{author}{\bibfnamefont{F.}~\bibnamefont{Denef}},
  \bibinfo{author}{\bibfnamefont{S.}~\bibnamefont{El-Showk}},
  \bibinfo{author}{\bibfnamefont{I.}~\bibnamefont{Messamah}}, \bibnamefont{and}
  \bibinfo{author}{\bibfnamefont{D.}~\bibnamefont{Van~den Bleeken}},
  \bibinfo{journal}{JHEP} \textbf{\bibinfo{volume}{11}}, \bibinfo{pages}{050}
  (\bibinfo{year}{2008}), \eprint{0802.2257}.

\bibitem[{\citenamefont{Brown and Henneaux}(1986)}]{Brown:1986ed}
\bibinfo{author}{\bibfnamefont{J.~D.} \bibnamefont{Brown}} \bibnamefont{and}
  \bibinfo{author}{\bibfnamefont{M.}~\bibnamefont{Henneaux}},
  \bibinfo{journal}{J. Math. Phys.} \textbf{\bibinfo{volume}{27}},
  \bibinfo{pages}{489} (\bibinfo{year}{1986}).

\bibitem[{\citenamefont{Azeyanagi et~al.}(2009)\citenamefont{Azeyanagi, Ogawa,
  and Terashima}}]{Azeyanagi:2008kb}
\bibinfo{author}{\bibfnamefont{T.}~\bibnamefont{Azeyanagi}},
  \bibinfo{author}{\bibfnamefont{N.}~\bibnamefont{Ogawa}}, \bibnamefont{and}
  \bibinfo{author}{\bibfnamefont{S.}~\bibnamefont{Terashima}},
  \bibinfo{journal}{JHEP} \textbf{\bibinfo{volume}{04}}, \bibinfo{pages}{061}
  (\bibinfo{year}{2009}), \eprint{0811.4177}.

\bibitem[{\citenamefont{Chen and Wang}(2010)}]{Chen:2009xja}
\bibinfo{author}{\bibfnamefont{C.-M.} \bibnamefont{Chen}} \bibnamefont{and}
  \bibinfo{author}{\bibfnamefont{J.~E.} \bibnamefont{Wang}},
  \bibinfo{journal}{Class. Quant. Grav.} \textbf{\bibinfo{volume}{27}},
  \bibinfo{pages}{075004} (\bibinfo{year}{2010}), \eprint{0901.0538}.

\bibitem[{\citenamefont{Amsel et~al.}(2009)\citenamefont{Amsel, Marolf, and
  Roberts}}]{Amsel:2009pu}
\bibinfo{author}{\bibfnamefont{A.~J.} \bibnamefont{Amsel}},
  \bibinfo{author}{\bibfnamefont{D.}~\bibnamefont{Marolf}}, \bibnamefont{and}
  \bibinfo{author}{\bibfnamefont{M.~M.} \bibnamefont{Roberts}},
  \bibinfo{journal}{JHEP} \textbf{\bibinfo{volume}{10}}, \bibinfo{pages}{021}
  (\bibinfo{year}{2009}), \eprint{0907.5023}.

\bibitem[{\citenamefont{Maldacena and Susskind}(1996)}]{Maldacena:1996ds}
\bibinfo{author}{\bibfnamefont{J.~M.} \bibnamefont{Maldacena}}
  \bibnamefont{and} \bibinfo{author}{\bibfnamefont{L.}~\bibnamefont{Susskind}},
  \bibinfo{journal}{Nucl. Phys.} \textbf{\bibinfo{volume}{B475}},
  \bibinfo{pages}{679} (\bibinfo{year}{1996}), \eprint{hep-th/9604042}.

\end{thebibliography}

\end{document}